# Predictability of catastrophic events; a new approach for SHM

D. Sornette[1] and C. Le Floc'h[2],

[1] LPMC, CNRS UMR6622 and Université des Sciences, 06100 Nice, France
and IGPP and ESS, UCLA, Los Angeles, CA90095, USA

[2] EADS Launch Vehicles, 33165 St Médard-en-Jalles, France

sornette@unice.fr   and   christian.le-floch@lanceurs.aeromatra.com

**Abstract**: We present a non-traditional general methodology for the prediction of breakdown of engineering structures, based on a global top-down monitoring of the health of the structure modelled by the theory of critical rupture. The new Acceptance Test Plan integrating the resulting Predictive Acoustic Emission procedure has been successfully applied to the prediction of the breakdown of composite over-wrapped high pressure vessels which are currently produced for the Ariane 5 launcher by EADS Launch Vehicles. This new approach should be an efficient tool more generally in Structural Health Monitoring service, especially for the prediction of the residual margins and lifetimes of heterogeneous structures which constitutes the main issue of SHM in civil infrastructures.

**Introduction**
The aim of Structural Health Monitoring (SHM) is to prevent catastrophes from occurring in service in almost all structures such as in transportation systems, public infrastructures and medical devices [1]. This is achieved with a combination of techniques, which include processing and modelling of the data obtained from sensors which are permanently installed on or within the structure during its life cycle. SHM allows one to improve significantly the safety of the structures and leads to cost reduction in operation and maintenance.

The challenge is to determine/predict the residual lifetime of a given structure, when subjected to a wide variety of environmental and a possibly complex history of mechanical loading conditions that can initiate damage and lead to failure [2].

The standard approach consists in detecting the elementary damage sources, in locating them and in estimating their impact regarding the capability of the structure to withstand the load. This analytical approach, based on mechanical modelling, requires a complex integrated system of sensors and involves sophisticated computations in order to combine the effect of multiple sources of damage. Indeed, this approach needs the identification of each damage location and of their specific mechanism, in order to be able to predict their evolution with time and eventually to analyse the impact of the combination of all of these mechanisms on the global failure of the structure.

Here, we would like to describe an alternative approach, which consists in considering the structure as a self-organising system under the influence of the external loading. In this complex system view, which could also be referred to as a systemic approach, a given structure is seen as developing a coherent large scale collective behaviour, which may eventually lead to a catastrophic failure. This synthetic system approach is inspired by the modelling of critical phenomena in statistical physics, which has been found to be relevant to a variety of different problems [3]. For material rupture in particular, this approach, which emphasizes and quantifies the collective and global behaviour rather than the details of all damage processes, has been found to be especially relevant for heterogeneous structures [4].

**When the heterogeneity of the material becomes an asset for prediction**
In strongly heterogeneous structures, the failure occurs as the culmination of the accumulation of many tiny damage events, that start uncorrelated but progressively self-organize by positive and negative feedbacks mediated by mechanical interactions. This organization, which can be viewed as a kind of correlated percolation, eventually culminates at the end of the life cycle by the global rupture of the structure. All the tiny damage events may belong to a variety of different mechanisms that compete and interact at many different scales.

This makes the problem very difficult to attack from an analytical standpoint, because such an approach needs a very good estimation of the balance between the different damage modes and a precise quantification of their interactions in order to construct a reliable bottom-up representation of the health of the structure. In other words, the conclusions drawn from the analytical approach are very sensitive to missed modes of damage and/or to an imprecise quantification of the cumulative damage and of the interactions between the different contributions to the damage processes.

In contrast, a top-down approach, based on the identification of global pertinent signatures of the overall health of the structure seems more appropriate and leads to more robust results. This claim is actually demonstrated by specific a priori calculations on solvable models of the failure of heterogeneous materials using mathematical tools borrowed from statistical physics and complex system theory and by the general approach of the renormalization group invented in the 1970s to deal with complex systems of many elements interacting at many scales (see [5] for an introduction).

Interestingly and somewhat paradoxically, this line of research has shown that the more heterogeneous is the material, the more warnings one gets! At a qualitative level, this was already noticed in the early 1960s by Mogi [6], a Japanese seismologist who led the Japanese research effort on earthquake prediction for several decades. This remark has recently been demonstrated in exactly soluble models and in numerical simulations systems constituted of assemblies of elastic-brittle elements [4]. The fracture processes thus strongly depend on the degree of heterogeneity of materials (quantified for instance by their distribution and topology of elastic modulii and their rupture thresholds). When the heterogeneity is large, then the rupture is more continuous and progressive. At the other extreme, a perfect crystal exhibits an abrupt failure process by a "one-nucleating-crack" mechanism with no significant damage precursors. Our approach does not apply to this class of homogeneous systems.

According to this model, rupture is the culmination of the progressive nucleation, growth and fusion of many tiny damage events. This is similar to a correlated percolation process in which the control parameter is progressively varied as a function of increasing strain, stress or simply time. The collective behaviour of the progressively self-organizing damage events is captured by an average deterministic power law at the macroscopic level, which is the well-known signature of a critical phenomenon [7]. The power law is characterized by:
- an absence of characteristic scale and of characteristic time,
- the existence of a critical point which is the mathematical expression of the global rupture.

In practice, if the structure is sufficiently heterogeneous that the critical rupture theory can be applied, the prediction of the failure load can in principle be obtained by fitting the experimental data with the theoretical power law function, whose parameters include the failure load.

**Case of an engineering structure: the composite over-wrapped high pressure vessel.**

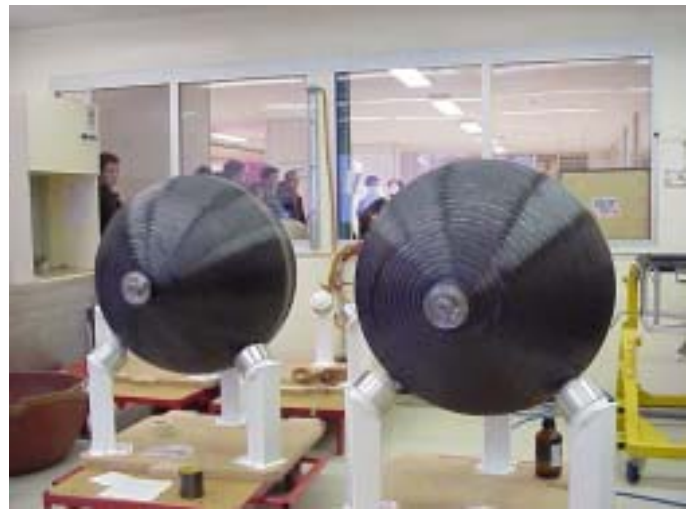

Figure 1: Ariane 5 composite high pressure tanks

Pressure tanks are vessels used for the storage of high pressure Helium gas which are used in the chemical propulsion system. For space applications, the tank must present a high performance, be lightweight and be designed to withstand severe launch and operational loads. To meet these common requirements, the high pressure vessel is made as a titanium liner, carbon fiber over-wrapped tank .

A key step of the industrial fabrication process, which will also be very important for the prediction, is the sizing pressure; a systematic pressurisation of the tank is performed immediately after construction to match the strains of the metallic and composite materials. This ensures a linear elastic behaviour during the operating life up to the proof pressure.

The real time monitoring of the structure during its first loading is performed with suitable sensors. In the present case, we use accelerometers, which record the acoustic emissions radiated by the tiny damage events within the composite carbon material during loading. By aggregating all acoustic emissions, we obtain a direct and robust measure of the global evolving state

of damage within the structure. This global measure of damage is represented by a power law

$$E = E_0/(P_R-P)^\alpha \quad (1)$$

where the cumulative acoustic emission energy (E) is represented as a function of the applied pressure (P) during the sizing pressure.

The signal processing consists simply in fitting the experimental data (E versus P) with the theoretical curve (1) and to identify the parameters ($E_0$, $\alpha$ and $P_R$), where $P_R$ is the pressure of rupture of the tank, $E_0$ a parameter of normalisation and $\alpha$ characterises the structure under loading [8,4]. Such power law (1) has been predicted analytically and verified by numerical simulations [9] as well as well-controlled laboratory experiments on analog electrical breakdown [10] and fiber-based materials [11].

In practice, the power law described in (1) is not operational. The real total acoustic energy released by the structure is much wilder and intermittent than predicted by the smooth and gently increasing power law (1). Indeed, the increasing damage of the structure is not occurring through a monotonous continuous process but is rather a succession of bursts and more or less quiescent phases, accelerating progressively and rather irregularly as the rupture is approached. When fitting such rough data with the formula (1), one gets an unreliable and inaccurate estimation of $P_R$ which is quite sensitive to the range used and to other details of the recording, making it unsuitable for a good prediction.

In order to alleviate this problem, one needs to extend the description of the critical rupture point beyond the simple power law (1) to capture the complex intermittent and fluctuating nature of the increasing damage. Consequently, the power law must be more "complex". It turns out that this play of word captures the natural generalization of formula (1) to take into account the physical processes of intermittent growth and fusion of micro-cracks [8,12,13]. Indeed, the most robust and relevant generalization of the simple power law (1) to account for these facts can be shown to amount to consider the exponent $\alpha$ as a complex number [14]. By taking the real part of the formula, this introduces so-called log-periodic oscillations decorating the basic power law (1), that is, by multiplying (1) by a cosine of the logarithm of the distance from the critical load $P_R$. This non-trivial generalization of the power law formula (1) allows one to have a much better estimation of the critical load $P_R$, because this critical load now appears both in the power law pre-factor but also in the cosine function describing the average intermittency of the damage process. Oscillations are well-known to help in signal processing because one can lock-in on the oscillations to improve the signal-over-noise ratio. A similar effect applies here and provides a good characterization of the growing damage and of the critical load from acoustic emissions recordings.

A first version of the software based on this physical modelling has been developed by XRS company, and has later been improved and extended by us. The validation of the physical modelling and the prediction procedure has been performed on eight Ariane 5 standard flight model vessels. A typical experiment consists in having six AE transducers located at equal distances on the tropics of Cancer and Capricorn of the spherical tank, in order to get the best monitoring and coverage of all important AE events occurring during the sizing pressure test. The set-up of the signal recording channels is optimised in order to get a good signal-to-noise ratio.

The AE recordings are used above 40Mpa, in order to remove the irrelevant contribution of AE due to the plastic deformation of the metallic liner. The AE are recorded for an applied pressure load up to Pmax= 68MPa corresponding to the maximum pressure level of the sizing pressure test.

|         | Pmax | prediction | burst | Delta% |
|---------|------|------------|-------|--------|
| D6 (SN8)  | 1*   | 1,0625     | 1,083 | 2      |
| D7(SN15)  | 0,85 | 1,075      | 1,07  | 1      |
| Q2(SN17)  | 0,85 | 1          | 0,99  | 1      |
| Q1(SN18)  | 0,85 | 1          | 0,99  | 1      |
| SN35      | 0,85 | 1,075      | 1,071 | 1      |
| SN41      | 0,85 | 1,075      | 1,062 | 1      |

Note: * burst test

Figure 2: Table summarizing the results on rupture forecasts obtained on full-scale models. The pressures are expressed in units of the theoretical calibrated pressure Pburst for rupture.

The predictions obtained by this procedure are very good, with an accuracy around 1% at a maximum pressure 15% below the real pressure load. The two last predictions reported in the two lines at the bottom of Figure 2 have been performed during blind tests and provide out-of-sample confirmation of the quality of the prediction method.

This methodology has been applied on 15 tanks in parallel with the already existing non-destructive investigation (NDI) methods (ultrasonics and X-rays computed tomography) during the production phase. This learning phase was needed before abandoning the traditional NDI in order to get confidence in the procedure. From the tank SN74 up to present, the new Acceptance Test Plan integrating this Predictive Acoustic Emission procedure has been applied on more than 10 new tanks.

**Conclusions**

We have presented a non-traditional general methodology for the scientific predictions of catastrophic events in heterogeneous structures based on the concepts and techniques of statistical and non-linear physics.

It has been successfully applied to the prediction of the breakdown of composite over-wrapped high pressure vessels which are currently produced for the Ariane 5 launcher by EADS Launch Vehicles.

We believe that this new approach should be an efficient tool in Structural Health Monitoring service, especially for the prediction of the residual margins and lifetimes of heterogeneous structures which constitutes the main issue of SHM in civil infrastructures.

Moreover, as they occur in various disciplines, catastrophes have similarities in complex systems both in the failure of standards models and the way that such systems evolve toward them. This methodology has been also applied successfully to problems as varied as stock market crashes, with potential for earthquakes [3].